\RequirePackage{fix-cm}
\documentclass[smallextended]{svjour3}       
\smartqed  
\usepackage{graphicx}
\journalname{Wireless Personal Communications}
\begin{document}

\title{Internet of Things - Applications and Challenges in Technology and Standardization
}
\titlerunning{Internet of Things}        
\author{Debasis Bandyopadhyay\and Jaydip Sen
}

\institute{Debasis Bandyopadhyay \at
					Innovation Labs, Tata Consultancy Services Ltd.
              Kolkata - 700091, INDIA\\
              Tel.: +91-33-6636989\\
              Fax: +91-33-66367502\\
              \email{debasis.b@tcs.com} 
           \and
           Jaydip Sen\at
					Innovation Labs, Tata Consultancy Services Ltd.
              	Kolkata-700091, INDIA\\
					Tel.: +91-33-6636 7137\\
					FAX: +91-33-6636 7502\\
					\email{jaydip.sen@tcs.com}
}
\date{Received: date / Accepted: date}

\maketitle

\begin{abstract}
The phrase \textit{Internet of Things} (IoT) heralds a vision of the future Internet where connecting physical things, from banknotes to bicycles, through a network will let them take an active part in the Internet, exchanging information about themselves and their surroundings. This will give immediate access to information about the physical world and the objects in it – leading to innovative services and increase in efficiency and productivity. This paper studies the state-of-the-art of IoT and presents the key technological drivers, potential applications, challenges and future research areas in the domain of IoT. IoT definitions from different perspective in academic and industry communities are also discussed and compared. Finally some major issues of future research in IoT are identified and discussed briefly. 
\end{abstract}

\section{Introduction}
\label{intro}
During the past few years, in the area of wireless communications and networking, a novel paradigm named the Internet of Things (IoT) which was first introduced by Kevin Ashton in the year 1998, has gained increasingly more attention in the academia and industry \cite {Santucci1}. By embedding short-range mobile transceivers into a wide array of additional gadgets and everyday items, enabling new forms of communication between people and things, and between things themselves, IoT would add a new dimension to the world of information and communication.

Unquestionably, the main strength of the IoT vision is the high impact it will have on several aspects of every-day life and behavior of potential users. From the point of view of a private user, the most obvious effects of the IoT will be visible in both working and domestic fields. In this context, assisted living, smart homes and offices, e-health, enhanced learning are only a few examples of possible application scenarios in which the new paradigm will play a leading role in the near future \cite {Atzori}. Similarly, from the perspective of business users, the most apparent consequences will be equally visible in fields such as automation and industrial manufacturing, logistics, business process management, intelligent transportation of people and goods.
 
However, many challenging issues still need to be addressed and both technological as well as social knots need to be united before the vision of IoT becomes a reality. The central issues are how to achieve full interoperability between interconnected devices, and how to provide them with a high degree of smartness by enabling their adaptation and autonomous behavior, while guaranteeing trust, security, and privacy of the users and their data \cite {Heuser}. Moreover, IoT will pose several new problems concerning issues related to efficient utilization of resources in low-powered resource constrained objects.

Several industrial, standardization and research bodies are currently involved in the activity of development of solutions to fulfill the technological requirements of IoT. The objective of this paper is to provide the reader a comprehensive discussion on the current state of the art of IoT, with particular focus on what have been done in the areas of protocol, algorithm and system design and development, and what are the future research and technology trends. 
 
The rest of the paper is organized as follows. Section~\ref{sec:1} presents the vision of IoT. Section~\ref{sec:2} discusses a generic layered architectural framework for IoT and various issues involved in different layers. Section~\ref{sec:3} presents the key technologies involved in IoT. Section~\ref{sec:4} presents some specific applications of IoT in various industry verticals. Section~\ref{sec:5} identifies some of the challenges in deploying the concept of IoT in the real. Section~\ref{sec:6} presents future research areas in the domain of IoT, and Section~\ref{sec:7} concludes the paper.

\section{Visions of Interent of Things}
\label{sec:1}
In the research communities, IoT has been defined from various different perspectives and hence numerous defintions for IoT exist in the literature. The reason for apparent fuzziness of the defintion stems from the fact that it is syntactically composed of two terms - Internet and things. The first one pushes towards a network oriented vision of IoT, while the second tends to move the focus on generic \textit{objects}. to be integrated into a common framework \cite{Atzori}. However, the terms 'Internet' and 'things', when put together assume a meaning which introduces a disruptive level of innovation into the ICT world. In fact, IoT semantically means a "world-wide network of interconnected objects uniquely addressable, based on standard communication protocols" \cite{Infso}. This implies a huge number of possibly heterogeneous objects involved in the process. In IoT, unique identification of objects and the representation and storing of exchanged information is the most challenging issue. This brings the third perspective of IoT - semantic perspective.
\begin{figure}
\includegraphics{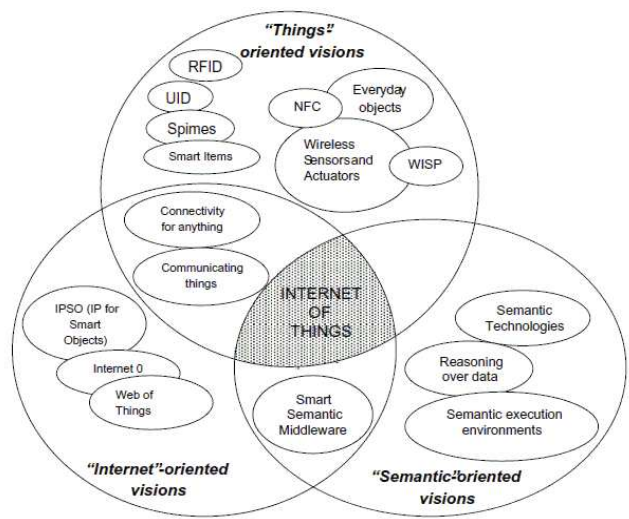}
\caption{Convergence of different visions of IoT}
\label{fig:1}  
\end{figure}
In Fig.~\ref{fig:1}, the main concepts, technologies and standards are highlighted and classified with reference to the three visions of IoT \cite{Atzori}. The diagram clearly depicts that IoT paradigm will lead to the convergence of the three visions of IoT.

From the persspective of \textit{things}, the focus of IoT is on how to integrate generic objects into a common framework and the \textit{things} under investigation are redio frequency identification (RFID) tags.The term IoT, in fact, is attributed to the Auto-ID labs \cite{Auto-ID}, a world-wide network of academic research laboratories in the field of networked RFID and emerging sensing technologies. These institutions, since their establishment, have focussed their efforts to design the architecture of IoT integrated with EPC global \cite{EPCglobal}. There efforts have been primarily towards development of the electronic product code (EPC) to support the use of RFID in the world-wide modern trading networks, and to create the industry-driven global standards for the EPC global Network. These standards are mainly designed to improve object visibility (i.e. the traceability of an object and the awareness of its status, current location etc.). While, this is an important step towards the deployment of IoT, it makes the scope of IoT much narrower. In a broader sense, IoT cannot be just a global EPC system in which the only objects are RFIDs. Similarly, unique/universal/ubiquitous identifier (UID) architecture defined in \cite{Sakamura} which attempts to develop middleware-based solutions for global visibility of objects also narrows down the scope of IoT. An IoT vision statement, which goes well beyond a mere rfid-centric approach, is proposed by CASAGARAS consortium \cite{CASAG}. The CASAGARAS consortium (i) proposes a vision of IoT as a global infrastructure which connects both virtual and physical generic objects and (ii) highlights the importance of including existing and evloving Internet and network developments in this vision. From this perspective, IoT becomes the natural enabling architecture for the deployment of independent federated services and applications, characterized by a high degree of autonomous data capture, event transfer, network connectivity and interoperability.

While the perspective of \textit{things} focuses on integrating generic objects into a common framework, the perspective of 'Internet' pushes towards a network-oriented definition. According to IPSO (IP for Smart Objects) alliance \cite{CASAG}, a forum formed in the year 2008, the IP stack is a light-weight protocol that already connects a large number of communicating devices and runs on battery-operated devices. This guarantees that IPhas all the qualities to make IoT a reality. It is likely that through an intelligent adaptation of IP and by incorporating IEEE 802.15.4 protocol into the IP architecture, and by adoption of 6LoWPAN \cite{Hui}, a large-sacle deployment of IoT will be reality.

As mentioned earlier in this section, semantic oriented IoT visions have also been proposed in the literature \cite{Toma}, \cite{Katasonov}, \cite{Wahlster}, \cite{Vazquez}. The idea behind this proposition is that the number of items involved in the future Internet is destined to become extremely high. Therefore, issues pertaining how to represent, store, interconnect, search, and organize information generated by the IoT will become very challenging. In this context, semantic technologies will play a key role. in fact, these technologies can exploit appropriate modeling solutions for \textit{things} description, reasoning over data generated by IoT, semantic execution environments and architectures that accomodate IoT requirements and scalable storing and communication infrastructure \cite{Toma}. 

A further vision correlated with the IoT is the so called \textit{web of things}\cite{Guinard}. Accoring to this vision of IoT, web standards are reused to connect and integrate into the web every-day-life objects that contain an embedded device or computer.

\section{Architecture of Internet of Things}
\label{sec:2}
\begin{figure}
\includegraphics{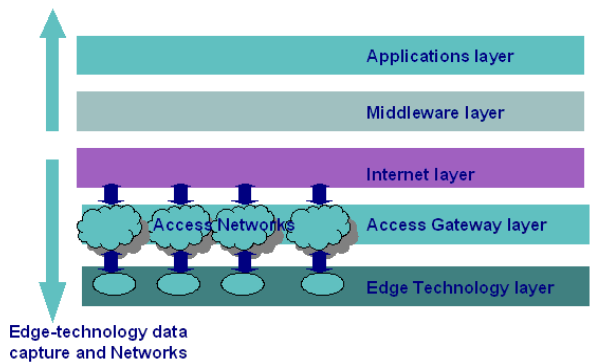}
\caption{Layered architecture of Internet of Things}
\label{fig:2}  
\end{figure}

Implementation of IoT is based on an architecture consisting of several layers: from the field data acquisition layer at the bottom to the application layer at the top. The layered architecture is to be designed in a way that can meet the requirements of various industries, enterprises, socities, institutes, goverments etc. Fig.~\ref{fig:2} presents a generic layered architecture for IoT [2].
The layered architecture has two distinct divisions with an Internet layer in between to serve the purpose of a common media for communication. The two lower layers contribute to data capturing while the two layers at the top is responsible for data utilization in applications. The functionalities of the various layers are discussed briefly in the following:
\begin{itemize}
	\item \textbf{Edge layer}: this hardware layer consists of sensor networks, embedded systems, RFID tags and readers or other soft sensors in different forms. These entities are the primary data sensors deployed in the field. Many of these hardware elements provide identification and information storage (e.g. RFID tags), information collection (e.g. sensor networks), information processing (e.g. embedded edge processors), communication, control and actuation.
	\item  \textbf{Access gateway layer}: the first stage of data handling happens at this layer. It takes care of message routing, publishing and subscribing and also performs cross platform communication, if required.
	\item \textbf{ Middleware layer}: this is one of the most critical layers that operates in bidirectional mode. It acts as an interface between the hardware layer at the bottom and the application layer at the top. It is responsible for critical functions such as device management and information management and also takes care of issues like data filtering, data aggregation, semantic analysis, access control, information discovery such as EPC (Electronic Product Code) innformation service and ONS (Object Naming Service). 
	\item \textbf{Application layer}: this layer at the top of the stack is responsible for delivery of various applications to different users in IoT. The applications can be from different industry verticals such as: manufacturing, logistics, retail, environment, public safety, healthcare, food and drug etc. With the increasing maturity of RFID technology, numerous applications are evolving which will be under the umbrella of IoT.
\end{itemize}
 \section{Key Technologies Involved in Internet of Things}
\label{sec:3}
IoT can only be realized by useful deployment of multiple technologies that covers in the domain of Hardware, Software and extremely robust applications around each domain of industries and operating sectors.
In this context, this Section will present the technology areas enabling the IoT and will identify the research and development challenges and outline a roadmap for future research activities to provide practical and reliable solutions. Some of the key technology areas that will enable IoT are: (i) identification technology, (ii) IoT architecture technology, (iii) communication technology, (iv) network technology, (v) network discovery technology, (vi) softwares and algorithms, (vii) hardware technology, (viii) data and signal processing technology, (ix) discovery and search engine technology, (x) relationship network management technology, (xi) power and energy storage technology, (xii) security and privacy technologies, and (xiii) standardization. These key technology enablers are discussed briefly in the following subsections.
\subsection{Identification technology}
\label{sec:3.1}
The function of identification is to map a unique identifier or UID (globally unique or unique within a particular scope), to an entity so as to make it without ambiguity identifiable and retrievable. UIDs may be built as a single quantity or out of a collection of attributes such that the combination of their values is unique. In the vision of IoT, things have a digital identity (described by unique identifiers), are identified with a digital name and the relationships among things can be specified in the digital domain. 
IoT deployment will require the development of new technologies that need to address the global ID schemes, identity management, identity encoding/encryption, authentication and repository management using identification and addressing schemes and the creation of global directory lookup services and discovery for IoT applications with various unique identifier schemes.
\subsection{IoT architecture technology}
\label{sec:3.2}
\begin{figure}
\includegraphics{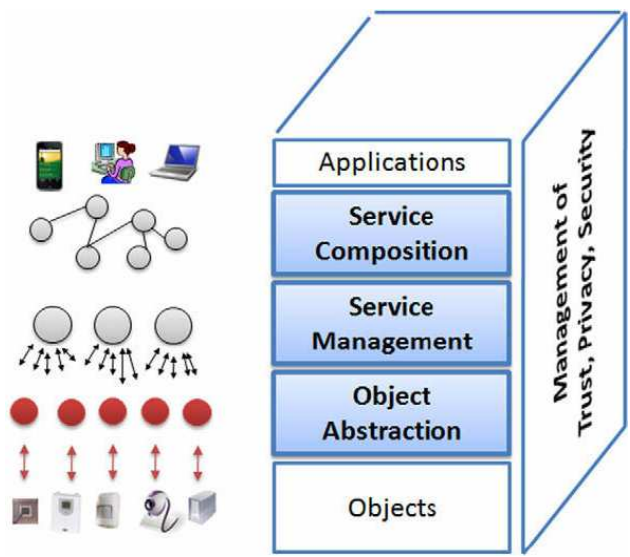}
\caption{SOA-based architecture for the IoT middleware}
\label{fig:3}  
\end{figure}

The middleware (a software layer interposed between the technological and application levels) architectures proposed in the last couple of years for IoT often follow the service oriented architecture (SOA) approach. The adoption of the SOA principles allows for decomposing complex and monolithic systems into applications consisting of ecosystem of simpler and well-defined components. The use of common interfaces and standard protocols gives a horizontal view of an enterprise system. Therefore, the development of business process of designing workflows of coordinated services, which eventually are associated with objects actions. An SOA approach also allows for software and hardware reuse, because it does not impose a specifc technology for service implementation \cite{Pasley}. Fig.~\ref{fig:3} presents a generic SOA-based architecture for the IoT middleware.
In the service oriented architecture (SoA), it becomes imperative for the providers and requestors to communicate meaningfully with each other despite the heterogeneous nature of the underlying information structures, business artifacts, and other documents. This requirement is termed as \textit{semantic interoperability}. Often technology is perceived to be the biggest impediment to effective collaboration and integration between requestors and providers; however, it is usually the problem of semantic interoperability which is the root cause. Semantic interoperability can be achieved between heterogeneous information systems (service providers and service requestors) in a multitude of ways. On one extreme, development of comprehensive shared information models can facilitate semantic interoperability among the participant applications and businesses. However, the problem with this approach is its rigidity, which translates to inflexibility. On the other extreme, semantic interoperability can be achieved by providing appropriate semantic mediators (translators) at each participant's end to facilitate the conversion to the information format which the participant understands. Most often systems use a combination of context independent shared information models coupled with context specific information specialization approaches to achieve semantic interoperability.

Scalability, modularity, extensibility and interoperability among heterogeneous \textit{things} and their environments are the key design requirements for IoT. Industry practitioners have suggested leveraging work in the semantic web to devise comprehensive and open ontologies to address the issue of semantic interoperability for dynamic binding based SOA for IoT application design and development.

\subsection{Communication technology}
\label{sec:3.3} 
The applications of IoT form an extensive design space with many dimensions that include several issues and parameters some of which are mentioned below.
\begin{itemize}
		\item Deployment – onetime, incremental or random.
		\item Mobility – occasional or continuous performed by either selected or all \textit{things} in the selected environment.
		\item Cost, size, resources, and energy – very much resource-constrained or unlimited resources.
		\item Heterogeneity – a single type of \textit{thing} or diverse sets of different properties and hierarchies.
		\item Communication modality – singel-hop or multi-hop communication.
		\item Infrastructure – different applications exclude, allow or require the use of fixed infrastructure.
		\item Network topology – single hop, star, multi-hop, mesh or multi-tier.
		\item Coverage- sparse, dense or redundant.
		\item Connectivity – continuous, occasional or sporadic.
		\item Network size – ranging from tens of nodes to thousands.
		\item Lifetime – few hours, several months to many years.
		\item QoS requirements - real-time constraints, tamper resistance, unobtrusiveness etc.
\end{itemize}
Such an extensive design space obviously makes IoT application development a complicated process. One approach may be is to make the design for the most restrictive point in the design space, e.g. minimum \textit{thing} capabilities, high mobility etc. However, often there is no such global minimum and it may be desirable to exploit the characteristics of the various points in the design space. This implies that no single hardware and software platform will be sufficient to support the whole design space. Complex and heterogeneous systems will be a natural requirement.
\subsection{Networking technology}
\label{sec:3.4} 
The IoT deployment requires developments of suitable network technology for  implementing the vision of IoT to reach out to objects in the physical world and to bring them into the Internet. Technologies like RFID, short-range wireless communication and sensor networks are means to achieve the network connectivity, while Internet protocol version 6 (IPv6), with its expanded address space, enables addressing, connecting and tracking things. 

In IoT paradigm, security, scalability, and cross platform compatibility between diverse networked systems will be essential requirements. In this context, the network technologies has to offer solutions that can offer the viability of connecting almost anything to the network at a reduced cost. The ubiquity of network access will also change the way information is processed. Today, IP provides end to end communication between devices without any requirement of an intermediate protocol translation gateway. Protocol gateways are inherently complex to design, manage, and deploy and with the end to end architecture of IP, there are no protocol translation gateways involved.

New scalable architectures designed specifically for the ubiquitous sensor networks communications will allow for networks of billions of devices. Improvements in techniques for secure and reliable wireless communication protocols will enable mission-critical applications for ubiquitous sensor networks based on wireless identifiable devices.
\subsection{Network discovery mechanisms}
\label{sec:3.5}
In the IoT paradigm, the networks will dynamically change and continuously evolve. Also, the \textit{things} will have varying degrees of autonomy. New \textit{things} will possibly be added and the network topologies will be changing fast. In this scenario, automated discovery mechanisms and mapping capabilities are essential for efficient network  and communication management. Without an automated discovery mechanism, it impossible to achieve a scalable and accurate network management capability. Moreover, an automated network discovery mechanism can dynamically assign roles to devices based on intelligent matching against pre-set templates and attributes, automatically deploy and start active, passive or performance monitors based on assigned roles and attributes, start, stop, manage and schedule the discovery process and make changes to any role or monitoring profile at any time or create new profiles as required.

Dynamic network discovery mechanisms enable interaction between devices that is not pre-configured and hard coded as far as the addresses or service end-points are concerned. Instead, they allow for dynamic, run-time configuration of connections, thereby enabling mobile devices to form collaborative groups and adapt to changing contexts. Examples for protocols for discovery on LAN level are WS-Discovery as a part of WSDD \cite{Oasis}, Bonjour \cite{Bonjour} and SSDP as a part of UPnP \cite{SSDP}.

Both passive and dynamic discovery mechanisms exist today and technologies are being developed to implement mechanisms real-time and dynamic discovery of network data. All discovery services must be based on authentication mechanisms to address privacy or security issues.
\subsection{Softwares and algorithms}
\label{sec:3.6}
One of the most promising micro operating systems for constrained devices is Contiki \cite{Contiki}. It provides a full IP stack (both IPv4 and IPv6), supports a local flash file system, and features a large development community and a comprehensive set of development tools. One of challenges in building IoT applications is how to design a common underlying software fabric for different environments and how to build a coherent application out of a large collection of diverse software modules. A substantial amount of research and development effort is currently focussed on service oriented computing for developing distributed and federated applications to support interoperable machine-to-machine and thing-to-thing interaction over a network. This is based on the Internet protocols, and on top of that, defines new protocols to describe and address the service instances. Service oriented computing loosely organizes the web services and makes it a virtual network.
\subsection{Hardware}
\label{sec:3.7} 
In the hardware front, research on nano-electronics devices is focussed on miniaturization, low cost and increased functionality in design of wireless identifiable systems. 

Silicon IC technology will be used for designing systems with increased functionalities and possessing enhanced non volatile memory for sensing and monitoring ambient parameters. Further research is needed in various areas such as: ultra-low power, low voltage and low leakage designs in submicron RF CMOS technologies, high-efficiency DC-DC power-management solutions, ultra low power, low voltage controllable nonvolatile memory, integration of RF MEMS and MEMS devices etc. The focus of research will be particularly on highly miniaturized integrated circuits that will include: (i) multi RF, adaptive and reconfigurable front ends, (ii) HF/UHF/SHF/EHF, (iii) memory- EEPROM/FRAM/Polymer, (iv) multi communication protocols, (v) digital processing, and (vi) security, including  tamper-resistance countermeasures, and technology to thwart side-channel attacks.
 
IoT will create new services and new business opportunities for system providers to service the communication demands of potentially tens of billions of devices in future. Following major trends are being observed in use of RFID tags. 

•	Use of ultra low cost tags having very limited features is observed. While the information is centralized on data servers managed by service operators, the value of information resides in the data management operations.
•	Use of low cost tags with enhanced features such as extra memory and sensing capabilities is also observed. The information is distributed both on centralized data servers and tags. The value resides in communication and data management, including processing of data into actionable information.
•	Use of smart fixed or mobile tags and embedded systems is also witnessed. More functionalities are brought into the tags bringing in local services. For such tags, information is centralized in the tags, while the value resides in the communication management to ensure security and effective synchronization with the network.

Smart devices with enhanced inter-device communication will lead to smart systems, which have high degrees of intelligence and autonomy enabling rapid deployment of IoT applications and creation of new services. 
\subsection{Data and signal processing technology}
\label{sec:3.8}
Industrial bodies from different domains have realized the utility of XML as the underlying language for standardization of business artifacts. Each industry vertical has come up with standardization bodies to develop XML standards for its own purpose. The primary objective of such effort is to develop a standardized way to express the contract, trust, process, workflow, message, and other data semantics in terms of XML nodes and attributes for the nodes. These XML vocabularies are then published as generalized \textit{document type definition} (DTD) or XML schema for consumption by members of that specific industry vertical. Since all members follow the same standardized DTD or schema, the semantic interoperability is achieved.

Initiatives such as International Standard for Metadata Registries (ISO/IEC 11179) and its implementation, e.g., the Universal Data Element Framework (UDEF) from OpenGroup aim to support semantic interoperability between structured data that is expressed using different schema and data dictionaries of vocabularies, by providing globally unique cross-reference identifiers for data elements that are semantically equivalent, even though they may have different names in different XML markup standards.

Finally, semantic web based standards from W3C like DAML (Darpa Agent Markup Language), RDF (Resource Description Framework) and OWL (Ontology Working Language) are useful in providing semantic foundations for dynamic situations involving dynamic discovery of businesses and services.

The intelligent decision-making algorithms will need to trigger activities not on the basis of a single event (such as an individual observation or sensor reading). Often these algorithms will have to consider correlation among events which may possibly require transformation of raw sensor data. Appropriate toolkits and frameworks already exist for complex event processing, such as ESPER and DROOLS - and are likely to play  useful roles in formulating machine-readable rules for determining the trigger sequences of events for a particular activity or process.
\subsection{Discovery and search engine technologies}
\label{sec:3.9} 
In IoT paradigm, information and services about \textit{things} will be fragmented across many entities and may be provided at class-level (i.e. common information and services for all instances of \textit{things} having the same class) or at serial-level (i.e. unique to an individual \textit{thing}), as well as being provided authoritatively by the creator of the \textit{thing} or contributed by other entities such as those who have interacted with an individual \textit{thing} in the past.

IoT will also require the development of lookup or referral services to link \textit{things} to information and services and to support secure access to information and services in a way that satisfies both the privacy of individuals and the confidentiality of business information. Such a matching between requesters and providers of information services can be based on trust relationships. As a smart \textit{thing} moves through the real world, it will encounter new environments, and both the smart thing and other agents that are monitoring the \textit{thing} will require lookup mechanisms in order to discover what capabilities are available within the local environment of the \textit{thing}. Such capabilities may include availability of sensors and actuators, network communication interfaces, facilities for computation and processing of data into information as well as facilities for onward transportation, handling, physical processing or alerting of a human operator about problems.
\subsection{Relationship network management technologies}
\label{sec:3.10}
IoT will require managing of networks that contain billons of heterogeneous \textit{things}, and where a wide variety of software, middleware and hardware devices exists. Network management technologies will have to addresses several important issues including, security, performance and reliability.

Network management involves managing distributed databases, repositories, auto polling of network devices, and real time graphical views of network topology changes and traffic. The network management service employs a variety of tools, applications, and devices to assist monitoring and maintaining the networks involved in IoT applications. Similar to the social network services that are flourishing today on the web, there would be a need for \textit{things} in the network to form relationships with each other. These relationships can be formal, such as membership within a federation, or they could be loosely based alliances brought upon by an incident or an event.
\subsection{Power and energy storage technologies}
\label{sec:3.11}
The autonomous \textit{things} operating in the IoT applications and performing either sensing or monitoring of the events need power and energy to perform the required job. Since the environments have wide variations depending on where and how the \textit{thing} is used, the power collection methods may vary, e.g., RF, solar, sound, vibration, heat, etc. In situations and locations where it is reasonable to have a large number of \textit{things} with sensing capabilities, use of mesh networks is a good proposition for increasing the communication and power efficiency by including the ability to forward transmissions from the closest \textit{thing}. The reader then only needs to be in range of the edge of the network.

Power and energy storage technologies are enablers for the deployment of IoT applications. These technologies have to provide high power-density energy generation and harvesting solutions which, when used with today's low power nano-electronics, will enable us to design self-powered intelligent sensor- based wireless identifiable device.
\subsection{Security and privacy technologies}
\label{sec:3.12} 
Two major issues in IoT are privacy of the humans and confidentiality of the business processes. Because of the scale of deployment, their mobility and often their relatively low complexity, the cloud of \textit{things} is hard to control. For ensuring confidentiality, a large number of standard encryption technologies exists for use. However, the main challenge is to make encryption algorithms faster and less energy-consuming. Moreover, an efficient key distribution scheme should be in place for using an encryption scheme. 

For small-scale systems, key distribution can happen in the factory or at the time of deployment, but for ad-hoc networks, novel key distribution schemes have only been proposed in recent years. For privacy, the situation is more serious; one of the reasons is the ignorance (regarding privacy) of the general public. Moreover, privacy-preserving technology is still in its infancy: the systems that do work are not designed for resource-restricted devices, and a holistic view on privacy is still to be developed (e.g., the view on privacy throughout one's life). The heterogeneity and mobility of 'things' in the IoT will add complexity to the situation. Also from a legal point of view, some issues remain far from clear and need legal interpretation; examples include the impact of location on privacy regulation, and the issue of data ownership in collaborative clouds of 'things' Network and data anonymity can provide a basis for privacy, but at the moment, these technologies are mainly supported by rather powerful equipment, in terms of computing power and bandwidth. A similar argument can be made for authentication of devices and establishing trust. 
\subsection{Standardization}
\label{sec:3.13} 
Standards should be designed to support a wide range of applications and address common requirements from a wide range of industry sectors as well as the needs of the environment, society and individual citizens. Through consensus processes involving multiple stakeholders, it will be possible to develop standardized semantic data models and ontologies, common interfaces and protocols, initially defined at an abstract level, then with example bindings to specific cross-platform, cross-language technologies such as XML, ASN.1, web services etc. The use of semantic ontologies and machine-readable codification should help to overcome ambiguities resulting from human error or differences and misinterpretation due to different human languages in different regions of the world, as well as assisting with cross-referencing to additional information available through other systems.
  
Standards are required for bidirectional communication and information exchange among things, their environment, their digital counterparts in the virtual cloud and entities that have an interest in monitoring, controlling or assisting the things. In addition, the design of standards for IoT needs to consider efficient and judicial use of energy and network capacity, as well as respecting other constraints such as those existing regulations that restrict permitted frequency bands and power levels for radio frequency communications. As IoT evolves, it may be necessary to review such regulatory constraints and investigate ways to ensure sufficient capacity for expansion, such as seeking additional radio spectrum allocation as it becomes available. A particular challenge in this regard is ensuring global interoperability particularly for things and devices that make use of radio spectrum. Historically, various bands of radio spectrum have been allocated for various purposes, such as broadcast communications (AM, FM, digital audio broadcasting, analogue terrestrial television, digital terrestrial television), mobile telephony, citizen-band radio, emergency services communications, wireless internet, short-range radio. Unfortunately, the frequency band allocations are not exactly harmonized across all regions of the world and some bands that are available for a particular purpose in one region are not available for the same purpose in another region, often because they are being used for a different purpose.

Re-allocation of radio spectrum is a slow process, involving government agencies, regulators and international bodies such as the International Telecommunications Union (ITU) as well as regional bodies such as the European Telecommunications Standards Institute (ETSI) or the Federal Communications Commission (FCC). Careful discussions are needed to minimize disruption to existing users of radio spectrum and to plan for future needs. In the meantime, many IoT devices using radio spectrum will need to be capable of using multiple protocols and multiple frequencies. An example of this is the ISO 18000-6C/EPCglobal UHF Gen2 standard, which is implemented using slightly different frequencies within the 860-960 MHz band, depending on the region of operation, as well as different power levels and different protocols (at least initially in Europe, where the Listen-Before-Talk protocol was required).

With regrad to the IoT paradigm at large, a very interesting standardization effort has currently started in ETSI \cite{ETSI}, which produces globally applicable ICT-related standards. Within ETSI, the Machine-to-Machine (M2M) Technical Committee was formed to conduct standardization activities relevant to M2M systems and sensor networks.M2M is a real leading paradigm towards IoT, but there has been very less standardization activities so far in this area. The goals of the ETSI M2M committee include: development and maintenance of an end-to-end architecture for M2M, strengthening the standardization efforts on M2M, including sensor network integration, naming, addressing, location,QoS, security, charging, management, application, and hardware interfaces \cite{Shelby}. 

As for the Internet Engineering Taskforce (IETF) activities related to the IoT, recently Ipv6 over Low Power Wireless PersonalArea Networks (6LoWPAN) IETF group has been formed \cite{Kushalnagar}. 6LoWPAN has defined a set of protocols that can be used to integrate sensor nodes into IPv6 networks. Core protocols for 6LoWPAN architecture have already been specified and some commercial procuts have been launched that implement this protocol suite.

Another working group in IETF' named \textit{routing over low power and lossy} (ROLL) networks, has recently produced RPL routing protocol draft. This will be the basis for routing over low power and lossy networks including 6LoWPAN. 

It is clear that an emerging idea is to consider the IoT standardization as an integral part of the future Internet definition and standardization process. This assertion has been made also by the cluster of European research and development projects on IoT (CERP-IoT). As presented in its report, the integration of different \textit{things} into wider networks, either mobile or fixed, will allow their interconnection with the future Internet \cite{Santucci2}.
\section{Applications of IoT}
\label{sec:4} 
The potentialities offered by the IoT make it possible to develop numerous applications based on it, of which only a few applications are currently deployed. In future, there will be intelligent applications for smarter homes and offices, smarter transportation systems, smarter hospitals, smarter enterprises and factories. In the following subsections, some of the important example applications of IoT are briefly discussed. 
\subsection{Aerospace and aviation industry}
\label{sec:4.1}  
IoT can help to improve safety and security of products and services by reliably identifying counterfeit products and elements. The aviation industry, for example, is vulnerable to the problem of \textit{suspected unapproved parts} (SUP). An SUP is an aircraft part that is not guaranteed to meet the requirements of an approved aircraft part (e.g., counterfeits, which do not conform to the strict quality constraints of the aviation industry). Thus, SUPs seriously violate the security standards of an aircraft. Aviation authorities report that at least 28 accidents or incidents in the United States have been caused by counterfeits \cite{CTV}. Apart from time-consuming material analyses, verifying the authenticity of aircraft parts can be performed by inspecting the accompanying documents, which can be easily forged. It is possible to solve this problem by introducing electronic pedigrees for certain categories of aircraft parts, which document their origin and safety-critical events during their lifecycle (e.g., modifications). By storing these pedigrees within a decentralized database as well as on RFID tags, which are securely attached to aircraft parts, an authentication (verification of digital signatures, comparison of the pedigree on RFID tags and within the database) of these parts can be performed prior to installing them in an aircraft. In this way, safety and operational reliability of aircrafts can be significantly improved.
\subsection{Automotive industry}
\label{sec:4.2} 
Advanced cars, trains, buses as well as bicycles are becoming equipped with advanced sensors, actuators with increased processing powers. Applications in the automotive industry include the use of \textit{smart things} to monitor and report various parameters from pressure in tyres to proximity of other vehicles. RFID technology has already been used to streamline vehicle production, improve logistics, increase quality control and improve customer services. The devices attached to the parts contain information related to the name of the manufacturer and when and where the product was made, its serial number, type, product code, and in some applications the precise location in the facility at that moment. RFID technology provides real-time data in the manufacturing processes, maintenance operations and offers new ways of managing recalls more effectively. \textit{Dedicated Short Range Communication} (DSRC) technology will possibly help in achieving higher bit rates and reducing interference with other equipment. \textit{Vehicle-to vehicle} (V2V) and  \textit{vehicle-to-infrastructure} (V2I) communications will significantly advance \textit{Intelligent Transportation Systems} (ITS) applications such as vehicle safety services and traffic management and will be fully integrated in the IoT infrastructure.
\subsection{Telecommunications industry}
\label{sec:4.3} 
IoT will create the possibility of merging of diverse telecommunication technologies and create new services. An illustrative example is the use of GSM, NFC (Near Field Communication), low power Bluetooth, WLAN, multi-hop networks, GPS and sensor networks together with SIM-card technology. In these types of applications the reader (i.e. tag) is a part of the mobile phone, and different applications share the SIM-card. NFC enables communications among objects in a simple and secure way just by having them close to each other. The mobile phone can therefore be used as a  NFC-reader and transmit the read data to a central server. When used in a mobile phone, the SIM-card plays an important role as storage for the NFC data and authentication credentials (like ticket numbers, credit card accounts, ID information etc). Things can join networks and facilitate peer-to-peer communication for specialized purposes or to increase robustness of communications channels and networks. Things can form ad-hoc peer-to-peer networks in disaster situations to keep the flow of vital information going in case of telecommunication infrastructure failures.
\subsection{Medical and healthcare industry}
\label{sec:4.4}
IoT will have many applications in the healthcare sector, with the possibility of using the cell phone with RFID-sensor capabilities as a platform for monitoring of medical parameters and drug delivery. The advantage gained is in prevention and easy monitoring of diseases, ad hoc diagnosis and providing prompt medical attention in cases of accidents. Implantable and addressable wireless devices can be used to store health records that can save a patient's life in emergency situations, especially for people with diabetes, cancer, coronary heart disease, stroke, chronic obstructive pulmonary disease, cognitive impairments, seizure disorders and Alzheimer's disease. Edible, biodegradable chips can be introduced into human body for guided actions. Paraplegic persons can have muscular stimuli delivered via an implanted smart thing-controlled electrical simulation system in order to restore movement functions.
\subsection{Independent living}
\label{sec:4.5}
IoT applications and services will have an important impact on independent living by providing support for an aging population by detecting the activities of daily living using wearable and ambient sensors, monitoring social interactions using wearable and ambient sensors, monitoring chronic disease using wearable vital signs sensors, and in body sensors. With emergence of pattern detection and machine learning algorithms, the \textit{things} in a patient's environment would be able to watch out and care for the patient. Things can learn regular routines and raise alerts or send out notifications in anomaly situations. These services can be merged with the medical technology services, mentioned in Section~\ref{sec:4.4}. 
\subsection{Pharmaceutical industry}
\label{sec:4.6}
For pharmaceutical products, security and safety is of utmost importance. In IoT paradigm, attaching smart labels to drugs, tracking them through the supply chain and monitoring their status with sensors has many potential benefits. For example, items requiring specific storage conditions, e.g. maintenance of a cool chain, can be continuously monitored and discarded if conditions were violated during transport. Drug tracking and e-pedigrees allow for the detection of counterfeit products and keep the supply chain free of fraudsters. Counterfeiting is a common practice in this area as illustrated in \cite{Kelesidis}, and it particularly affects the developing countries. The smart labels on the drugs can also directly benefit patients, e.g. by enabling storing of the package insert, informing consumers of dosages and expiration dates, and assuring the authenticity of the medication. In conjunction with a smart medicine cabinet that reads information transmitted by the drug labels, patients can be reminded to take their medicine at appropriate intervals and patient compliance can be monitored. 
\subsection{Retail, logistics and supply chain management}
\label{sec:4.7} 
IoT can provide several advantages in retail and supply chain management (SCM) operations. For example, with RFID-equipped items and smart shelves that track the present items in real time, a retailer can optimize many applications \cite{Hardgrave}. For example, he can make automatic checking of goods receipt, real time monitoring of stocks, tracking out-of-stocks or the detection of shoplifting. IoT can provide a large savings potential in a retail store, since it has been found that 3.9\% of sales loss happens worldwide when shelves go empty and customers return with getting the desired products \cite{Gruen}. Furthermore, IoT can help making the data from the retail store available for optimizing the logistics of the whole supply chain. If manufacturers know the stock and sales data from retailers, they can produce and ship the right quantities of products, thus avoiding the situation of over-production or underproduction. The logistic processes from supply chains in many industry sectors can benefit from exchanging of RFID data. Moreover, environmental issues can be better tackled. The carbon footprint of logistics and supply chain processes can be optimized based on the availability of dynamic and fine-grained data collected in the real world directly by some of the \textit{things} of IoT, such as trucks, pallets, individual product items etc. In the shops, IoT can offer many applications like guidance in the shop according to a pre-selected shopping list, fast payment solutions like automatically check-out using biometrics, detection of potential allergen in a given product, personalized marketing, verification of the cool chain, etc. Commercial buildings will also benefit from smart building functionalities.
\subsection{Manufacturing industry}
\label{sec:4.8} 
By linking items with information technology, either through embedded smart devices or through the use of unique identifiers and data carriers that can interact with an intelligent supporting network infrastructure and information systems, production processes can be optimized and the entire lifecycle of objects, from production to disposal can be monitored. By tagging items and containers, greater transparency can be gained about the status of the shop floor, the location and disposition of lots, and the status of production machines. The fine grained information serves as input data for refined production schedules and improved logistics. Self-organizing and intelligent manufacturing solutions can be designed around identifiable items.
\subsection{Process industry}
\label{sec:4.9} 
In many plants of the oil and gas industry, scalable architectures are being used that consider possibilities for plug-and-play new ID methods combined with sensing/actuating integrated with the IoT infrastructure and integrate the wireless monitoring of petroleum personnel in critical onshore and offshore operations, container tracking, tracking of drill string components pipes, monitoring and managing of fixed equipment etc. A review of high-impact accidents in the chemical and petrochemical sectors in the UK \cite{Fewtrell} has observed some common features in these disasters, such as lack of understanding as well as poor management of storage, process, and chemical segregation. IoT can help in reducing the number of accidents in the oil and gas industry by equipping the containers of hazardous chemicals with intelligent wireless sensor nodes.  
\subsection{Environment monitoring}
\label{sec:4.10}
Utilization of wireless identifiable devices and other IoT technologies in green applications and environmental conservation are one of the most promising market segments in the future. There will be an increased usage of wireless identifiable devices in environmentally friendly programs worldwide.
\subsection{Transportation industry}
\label{sec:4.11}
IoT offers solutions for fare collection and toll systems, screening of passengers and bags boarding commercial carriers and the goods moved by the international cargo system that support the security policies of the governments and the transportation industry, to meet the increasing demand for security in the globe. Monitoring traffic jams through cell phones of the users and deployment of \textit{intelligent transport systems} (ITS) will make the transportation of goods and people more efficient. Transportation companies would become more efficient in packing containers since the containers can self- scan and weigh themselves. Use of IoT technologies for managing passenger luggage in airports and airline operations will enable automated tracking and sorting, increased per-bag read rates, and increased security.
\subsection{Agriculture and breeding}
\label{sec:4.12}
The regulations for traceability of agricultural animals and their movements require the use of technologies like IoT, making possible the real time detection of animals, for example during outbreaks of contagious disease. Moreover, in many cases, countries give subsidies depending on the number of animals in a herd and other requirements, to farms with cattle, sheep, and goats. As the determination of the number is difficult, there is always the possibility of frauds. Good identification systems can help minimize this fraud. Therefore, with the application of identification systems, animal diseases can be controlled, surveyed, and prevented. Official identification of animals in national, intra community, and international commerce is already in place, while at the same time, identification of livestock that are vaccinated or tested under official disease control or eradication is also possible. Blood and tissue specimens can be accurately identified, and the health status of herds, regions, and countries can be certified by using IoT. With the Internet of Things, single farmers may be able to deliver the crops directly to the consumers not only in a small region like in direct marketing or shops but in a wider area. This will change the whole supply chain which is mainly in the hand of large companies, now, but can change to a more direct, shorter chain between producers and consumers. 
\subsection{Media, entertainmnet industry}
\label{sec:4.13}
Deployment of IoT technologies will enable ad hoc news gathering based on locations of the users. The news gathering could happen by querying IoT, to see which multi-media-capable devices are present at a certain location, and sending them a (financial) offer to collect multimedia footage about a certain event. Near field communication tags can be attached to posters for providing more information by connecting the reader to an URI address that contains detailed information related to the poster. 
\subsection{Insurance industry}
\label{sec:4.14}
Often the introduction of IoT technology is perceived as a grave invasion on privacy of individuals. However, sometimes people are willing to trade privacy for a better service or a monetary benefit. One example is car insurance. If insurance clients are willing to accept electronic recorders in their car, which are able to record acceleration, speed, and other parameters, and communicate this information to their insurer, they are likely to get a cheaper rate or premium \cite{Coroama}. The insurer can save money by being involved in a very early stage of an impending accident and can trigger the most economic actions. A part of the savings can be given to the customers through discounts on insurance premiums. The same applies for other assets such as buildings, machinery, etc., that are equipped with IoT technology. In these cases the technology mostly helps in preventing large-scale maintenance operations or allows for much cheaper predictive maintenance before an incident occurs.
\subsection{Recycling}
\label{sec:4.15}              
IoT and wireless technologies can be used to advance the efficiency and effectiveness of numerous important city and national environmental programs, including the monitoring of vehicle emissions to help supervise air quality, the collection of recyclable materials, the reuse of packaging resources and electronic parts, and the disposal of electronic waste (RFID used to identify electronic subcomponents of PCs, mobile phones, and other consumer electronics products to increase the reuse of these parts and reduce e-waste). RFID continues to provide greater visibility into the supply chain by helping companies more efficiently track and manage inventories, thereby reducing unnecessary transportation requirements and fuel usage.
\section{Challenges and Open Issues}
\label{sec:5}
The workflows in analysed enterprise environment, home, office and other smart spaces in the future will be characterized by cross organization interaction, requiring the operation of highly dynamic and ad-ho relationships. At present, only a very limited ICT support is available, and the following key challenges exist.
  
(i) \textbf{Network Foundation} - limitations of the current Internet architecture in terms of mobility, availability, manageability and scalability are some of the major barriers to IoT.

(ii) \textbf{Security, Privacy and Trust} - in the domain of security the challenges are: (a) securing the architecture of IOT - security to be ensured at design time and execution time, (b) proactive identification and protection of IOT from arbitrary attacks (e.g. DoS and DDoS attacks) and abuse, and (c) proactive identification and protection of IOT from malicious software. 
In the domain of user privacy, the specific challenges are: (a) control over personal information (data privacy) and control over individual's physical location and movement (location privacy), (b) need for privacy enhancement technologies and relevant protection laws, and (c) standards, methodologies and tools for identity management of users and objects. 
In the domain of trust, some of the specific challenges are: (a) Need for easy and natural exchange of critical, protected and sensitive data  - e.g. smart objects will communicate on behalf of users / organizations with services they can trust, and (b) trust has to be a part of the design of IoT and must be built in.

(iii) \textbf{Managing heterogeneity} - managing heterogeneous applications, environments and devices constitute a major challenge.
 
In addition to the above major challenges, some of the other challenges are: (a) managing large amount of information and mining large volume of data to provide useful services, (b) designing an efficient architecture for sensor networking and storage, (iii) designing mechanisms for sensor data discovery, (iv) designing sensor data communication protocols - senor data query, publish/subscribe mechanisms, (v) developing sensor data stream processing mechanisms, and (vi) sensor data mining - correlation, aggregation filtering techniques design.  Finally, standardizing heterogeneous technologies, devices, application interfaces etc. will also be a major challenge.
\section{Future Research Areas}
\label{sec:6} 
There are several areas in which further research is needed for making deployment of the concept of IoT reliable, robust and efficient. Some of the areas are identified in the following.

In identification technology domain, further research is needed in development of new technologies that address the global ID schemes, identity management, identity encoding/ encryption, pseudonimity, revocable anonymity, authentication of parties, repository management using identification, authentication and addressing schemes and the creation of global directory lookup services and discovery services for IoT applications with various identifier schemes.
In architecture design domain, some of the issues that need attention are: design of distributed open architecture with end-to-end characteristics, interoperability of heterogeneous systems, neutral access, clear layering and resilience to physical network disruption, decentralized autonomic architectures based on peering of nodes etc.
 
In communication protocol domain, the issues that need to be addressed are : design of energy efficient communication by multi frequency protocol, communication spectrum and frequency allocation, software defined radios to remove the needs for hardware upgrades for new protocols, and design of high performance, scalable algorithms and protocols.
 
In network technology domain further research is needed on network on chip technology considering on chip communication architectures for dynamic configurations design time parameterized architecture with a dynamic routing scheme and a variable number of allowed virtual connections at each output. In addition, power-aware network design that turns on and off the links in response to burst and dips of traffic on demand, scalable communication infrastructures design on chip to dynamically support the communication among circuit modules based on varying workloads and /or changing constraints are some of the important research issues.
\section{Conclusions}
\label{sec:7} 
When we look at today's state of the art technologies, we get a clear indication of how the IoT will be implemented on a universal level in the coming years. We also get an indication of the important aspects that need to be further studied and developed for making large-scale deployment of IoT a reality. It is observed that an urgent need exists for significant work in the area of governance of IoT. Without a standardized approach it is likely that a proliferation of architectures, identification schemes, protocols and frequencies will happen parallelly, each one targetted for a particular and specific use. This will inevitably lead to a fragmentation of the IoT, which could hamper its popularity and become a major obstacle in its roll out. Interoperability is a necessity, and inter-tag communication is a pre-condition in order for the adoption of IoT to be wide-spread.

In the coming years, technologies necessary to achieve the ubiquitous network society are expected to enter the stage of maturity. As the RFID applications find more acceptability, a vast amount of objects will be addressable, and could be connected to IP-based networks, to constitute the very first wave of the IoT. There will be two major challenges in order to guarantee seamless network access: the first issue relates to the fact that today different networks coexist; the other issue is related to the sheer size of the IoT. The current IT industry has no experience in developing a system in which hundreds of millions of objects are connected to IP networks. Other current issues, such as address restriction, automatic address setup, security functions such as authentication and encryption, and multicast functions to deliver voice and video signals efficiently will probably be overcome by ongoing technological developments.
Another very important aspect that needs to be addressed at this early stage is the one related to legislation. Various consumer groups in the Europe and the USA have expressed strong concerns about the numerous possibilities for this technology to be misused. A clear legislative framework ensuring the right of privacy and security level for all users must therefore be implemented. A sustained information campaign highlighting the benefits of this technology to the society at large must also be organized. 

Traditionally, the retail and logistics industry require very low cost tags with limited features; such as an ID number and some extra user memory area, while other applications and industries will require tags that will contain much higher quantity of data and will be more interactive and intelligent. \textit{Data}, in this context, may be seen as an \textit{object}, and under this vision a tag carries not only its own characteristics but also the operations it can handle. The amount of intelligence that the objects in IoT will need to have, and in which situations this intelligence is distributed or centralized becomes a key factor of development. As the IQ of the \textit{things} will grow, the pace of the development and study of behavioral requirements of these objects will also become more prevalent in order to ensure that these objects can co-exist in seamless and non-hostile environments. These types of tags will contain features ranging from sensors and actuators and will interact with the environment in which they are placed. One such example could be an interactive device placed in the human body with the scope of delivering the right medicine at the right place at the right time. In this context of higher mobility, portability and intelligence, two trends will influence the future development of smart systems: the increased use of \textit{embedded intelligence}, and the \textit{networking of embedded intelligence}. Another interesting paradigm which is emerging in the Internet of the Fututre context is the so called Web Squarred, which is an evaolution of the Web 2.0. It is aimed at integrating web and sensing technologies \cite{O'Reilly} together so as to enrich the content provided to the users. This is obtained by taking into account the information about the user context collected by the sensors (microphone, camera, GPS, etc.) deployed in the user terminals. Web Squarred can be considered as one of the applications running over the IoT like the web is today an importnat application running over the Internet.
 
This paper surveyed some of the most important aspects of IoT with particular focus on what is being done and what are the issues that require further research. While the current technologies make the concept of IoT feasible, a large number of challenges lie ahead for making the a large-scale real-world deployment of IoT applications. In the next few years, addressing these challenges will be a powerful driving force for networking and communication research in both industrial and academic laboratories.


\begin{thebibliography}{}

\bibitem{Santucci1} G. Santucci. From Internet to Data to Internet of Things. Proceedings of the International Conference on Future Trends of the Internet. (2009).
\bibitem{Atzori} L. Atzori, A. Lera, and G. Morabito. The Internet of Things: A Survey. Computer Networks 54(15), 2787-2805.  (2010).
\bibitem{Heuser} Lutz Heuser, Zoltan Nochta, Nina-Cathrin Trunk. ICT Shaping the World: A Scientific View. ETSI, WILEY Publication.(2008).
\bibitem{Infso} INFSO D.4 Networked Enterprise and RFID INFSO G.2 Micro and Nanosystems. In: Co-operation with the Working Group RFID of the ETP EPOSS, Internet of Things in 2020, Roadmap for the Future, Version 1.1, May 27. (2008). 
\bibitem{Auto-ID} Auto-Id Labs. Url: http://www.autoidlabs.org
\bibitem{EPCglobal} The EPCglobal Architecture Framework. EPCglobal Final Version 1.3, Approved 19 March, 2009. Url: http://www.epcglobalinc.org. (2009).
\bibitem{Sakamura} K. Sakamura. Challenges in the Age of Ubiquitous Computing: A Case Study of T-Engine - An Open Development Platform for Embedded Systems. In: Proceedings of ICSE'06, Shanghai, China. (2006).
\bibitem{CASAG} A. Dunkels, J.P. Vasseur. IP for Smart Objects, Internet Protocol for Smart Objects (ISO) Alliance, White Paper \#1. Url: http://www.ispo-alliance.org. (2009).
\bibitem{Hui} J. Hui, D. Culler, S. Chakrabarti. 6LoWPAN: Incorporating IEEE 802.15.4 into IP Architecture- Internet Protocol for Smart Objects (IPSO) Alliance, White Paper \# 3. Url: http://www.ispo-alliance.org. (2009).
\bibitem{Toma} I. Toma, E. Simperl, G. Hench. A Joint Roadmap for Semantic Technologies and the Internet of Things. In: Proceedings of the 3rd STI Roadmapping Workshop, Crete, Greece. (2009).
\bibitem{Katasonov} A. Katasonov, O. Kaykova, O. Khriyenko, S. Nikitin, V. Terziyan. Smart Semantic Middleware for the Internet of Things. In: Proceedings of the 5th International Conference on Informatics in Control,Automation and Robotics, Funchal, Mederia, Portugal. (2008).
\bibitem{Wahlster} W. Wahlster. Web 3.0: Semantic Technologies for the Internet of Services and of Things.Lecture at the 2008 Dresden Future Forum. (2008).
\bibitem{Vazquez} I. Vazquez. Social Devices: Semantic Technology for the Internet of Things. Week@ESI, Zamudio, Spain.(2009).
\bibitem{Guinard} D. Guinard, T. Vlad. Towards the Web of Things: Web Mashups for Embedded Devices. In: Proceedings of the International World Wide Web Conference (WWW 2009), Madrid, Spain. (2009).
\bibitem{Pasley} J. Pasley. How BPEL and SOA are Changing Web Services  Development. IEEE Internet Computing 9(3), 65-67. (2005). 
\bibitem{Oasis}OASIS Standard: Web-Services Dynamic Discovery,version 1.1. Url:http://docs.oasis-open.org/ws-dd/discovery/1.1/wsdd-discovery-1.1-spec.html
\bibitem{Bonjour}Bonjour Protocol Specification. 
Url: http://developer.apple.com/networking/bonjour
\bibitem{SSDP} Simple Service Discovery Protocol/1.0. Operating without an Arbiter. Url: http://quimby.gnus.org/internet-drafts/draft-cai-ssdp-v1-03.txt.
\bibitem{Contiki} Contiki: The Operating System for Connecting Next Billion Devices- the Internet of Things. Url: http://www.sics.se/contiki
\bibitem{ETSI} European Telecommunication Telecommunications Standards Institute. Url: http://www.etsi.org.
\bibitem{Shelby} Z. Shelby. ETSI M2M Standardization, March, 2009. 
Url: http://zachshelby.org.
\bibitem{Kushalnagar} N. Kushalnagar, G. Montenegro, C. Schumacher. IPv6 over Lo-Power Wireless Personal Area Networks (6LoWPANs): Overview,Assumptions, Problem Statement, and Goals, IETF RFC 4919, August 2009.
\bibitem{Santucci2} G.Santucci. Internet of the Future and Internet of Things: What is at Stake and How are We Getting Prepared for Them? In: eMatch'99- Future Internet Workshop, Oslo, Norway,September, 2009.
\bibitem{CTV} CTV Deadly Fakes- CTV News. Url:http://www.ctv.ca/servlet/ArticleNews/story/CTVNews/20020306/ctvnews848463
\bibitem{Kelesidis} T. Kelesidis, I. Kelesidis, P. Rafailidis, and M. Falagas, "Counterfeit or substandard  antimicrobial drugs: a review of the scientific drugs: a review of the scientific evidence", Journal of Antimicrobial Chemotherapy, 60(2): 214 - 236, August 2007.
\bibitem {Hardgrave} B.C. Hardgrave, M. Waller, and R. Miller, "RFID's impact on out of stocks: a sales velocity analysis", Research Report from the University of Arkansas, 2006.
\bibitem {Gruen} T. W. Gruen, D.S. Corsten, and S. Bharadwaj, "Retail out of stocks", Technical Report, 2002.
\bibitem{Fewtrell} P. Fewtrell, I.L Hirst, "A review of high-cost chemical/petrochemical accidents since Flixborough 1974", in: Loss Prevention Bulletin (1998), April, No. 140. URL: http://www.hse.gov.uk/comah/lossprev.pdf.
\bibitem{Coroama} V. Coroama,"The smart tachograph - individual accounting of traffic costs and its implications", Proceedings of Pervasive 2006, pp. 135 - 152, Dublin, Ireland, May 07 - 10, 2006.
\bibitem{O'Reilly} T. O'Reilly,J. Pahlka. The Web Squarred Era. Forbes, September, 2009. (2009).

\end{thebibliography}
\end{document}